\documentclass{PoS}
\usepackage{graphicx}
\graphicspath{{figs/}}
\newcommand{\<}{\langle}
\renewcommand{\>}{\rangle}
\newcommand{\beq}{\begin{equation}}
\newcommand{\eeq}{\end{equation}}
\newcommand{\beqn}{\begin{eqnarray}}
\newcommand{\eeqn}{\end{eqnarray}}
\newcommand{\D}{\Delta}
\def\simge{\mathrel{\rlap{\raise 0.511ex \hbox{$>$}}{\lower 0.511ex \hbox{$\sim$}}}}
\def\simle{\mathrel{\rlap{\raise 0.511ex \hbox{$<$}}{\lower 0.511ex \hbox{$\sim$}}}} 
\def\sla#1{\setbox0=\hbox{$#1$}\dimen0=\wd0                    
      \setbox1=\hbox{/} \dimen1=\wd1 \ifdim\dimen0>\dimen1
      \rlap{\hbox to \dimen0{\hfil/\hfil}} #1                        \else
      \rlap{\hbox to \dimen1{\hfil$#1$\hfil}}
      /   \fi} 

\newcommand{\nn}{\nonumber}

\newcommand{\ov}{\overline}

\newcommand{\ra}{\rightarrow}
\newcommand{\mc}{\mathcal}
\newcommand{\FA}{{\sf FeynArts}}
\newcommand{\Math}{{\sf Mathematica}}

\newcommand{\be}{\beta}
\newcommand{\ga}{\gamma}

\newcommand{\de}{\delta}

\newcommand{\ie}{\emph{i.e.} }
\newcommand{\eg}{\emph{e.g.} }

\PoS{PoS(HEP2005)221}

\title{NLO calculation of the $\D F = 2$ Hamiltonians in the MSSM and 
phenomenological analysis of the $B - \ov{B}$ mixing 
\footnote{Preprint: {\tt Rome 1420/05}, {\tt ~RM3-TH/05-12}}}

\ShortTitle{NLO calculation of the $\D F = 2$ Hamiltonians in the MSSM 
and analysis of the $B-\ov{B}$ mixing~~~}

\author{M.~Ciuchini\,$^a$, E.~Franco\,$^b$,
\speaker{D.~Guadagnoli\,}$^{~b}$, 
V.~Lubicz\,$^a$, V.~Porretti\,$^a$, L.~Silvestrini~$^b$\\\vspace{0.3cm}\\
$^a~$Dipartimento di Fisica, Universit\`a di Roma Tre and INFN, Sezione di Roma Tre,\\
Via della Vasca Navale 84, I-00146 Rome, Italy\\
$^b~$Dipartimento di Fisica, Universit\`a di Roma ``La Sapienza'' and INFN,
Sezione di Roma,\\
P.le A.~Moro 2, I-00185 Rome, Italy\\\vspace{0.3cm}\\
E-mail: \email{diego.guadagnoli@roma1.infn.it}
}

\abstract{We present the NLO corrections to the Wilson coefficients 
of the $\Delta F = 2$ Hamiltonians in the strong interacting sector of the MSSM, 
responsible for neutral meson oscillations.
Such corrections, combined with the NLO anomalous dimension matrix for 
the $\Delta F = 2$ operators and the corresponding hadronic matrix elements
calculated on the lattice, allow for the first time a full NLO phenomenological 
analysis of the observables related to meson oscillations. 
We present preliminary results in the $B_d-\ov B_d$ sector.}

\FullConference{International Europhysics Conference on High Energy Physics\\
July 21st - 27th 2005\\
Lisboa, Portugal}

\begin{document}

\section{Introduction}\label{sec:intro}
Observables related to meson oscillations represent among the best measured
quantities in particle physics. This is true particularly for the $K-\ov{K}$ and
the $B_d - \ov{B}_d$ cases, where experiments at $K$- and $B$-factories have
pushed precisions to the permil level and new quantities in the $D$ and $B_s$ 
sectors promise to add up thanks to now-running and forthcoming experiments.
Experimental commitment is related to the special interest 
meson oscillations have on the theoretical side: they constitute eminent 
sources of so-called flavor changing neutral currents (FCNC's) possibly accompanied 
by a sizeable violation of the CP symmetry.
FCNC's and CP violation are in turn the effects that most naturally (and
copiously) occur in basically all extensions of the SM, and in particular in
SUSY, that, notwithstanding the success of the SM, represents the ``standard
way'' beyond it \cite{Alta}. The most popular low-energy realization of SUSY 
is provided by the MSSM, that however features `by default' a huge space of parameters. 
Placing high precision constraints on the latter thus becomes of capital
importance, considering that it can provide key handles on possibly ungrasped
symmetries. To this aim, meson oscillations constitute privileged observables,
recalling their experimental and theoretical `virtues', as mentioned above.

The amplitudes for meson oscillations can be accessed in the Effective
Hamiltonian (EH) approach, through the so-called $\D F = 2$ EH's, reading as
\beqn
\mc{H}_{\rm eff}^{\D F = 2} = \sum_i \, C_i \, Q_i~,
\label{Heff}
\eeqn
where the $Q_i$ are local operators of increasing mass dimension and the $C_i$
the corresponding `couplings' (Wilson coefficients). In the MSSM and restricting
to the $\D B = 2$ case ($B_d - \ov{B}_d$) a suitable basis for the $Q_i$ up to
dimension 6 is given by \cite{GGMS} ($P_{R,L} = (1 \pm \ga_5)/2$)
\beqn
&&Q_1 ~=~ 
(\ov{\psi}_b^i \gamma^\mu_L \psi_d^i)~(\ov{\psi}_b^j \gamma_{\mu L}
\psi_d^j)~,~~~~
Q_2 ~=~ 
(\ov{\psi}_b^i P_L \psi_d^i)~(\ov{\psi}_b^j P_L \psi_d^j) ~,~~~~
Q_3 ~=~ 
(\ov{\psi}_b^i P_L \psi_d^j)~(\ov{\psi}_b^j P_L \psi_d^i) ~,\nn \\
&&Q_4 ~=~ 
(\ov{\psi}_b^i P_L \psi_d^i)~(\ov{\psi}_b^j P_R \psi_d^j) ~,~~~~~~
Q_5 ~=~ 
(\ov{\psi}_b^i P_L \psi_d^j)~(\ov{\psi}_b^j P_R \psi_d^i) ~,
\label{basis}
\eeqn
plus $\tilde{Q}_{1,2,3}$ obtained from $Q_{1,2,3}$ with the substitution 
$L \ra R$. In eq. (\ref{basis}) indices $i,\,j$ denote color.

The $C_i$ in eq. (\ref{Heff}) are calculated by requiring that the amplitude
$\mc A$ for meson oscillation, evaluated in the full (\ie in the MSSM) and in the 
effective theory (\ref{Heff}) be the same below a high-energy threshold $M_{\rm SUSY}$. 
Since the $C_i$ encode the short-distance `part' of the process, the equality 
$\mc A_{\rm full} = \mc A_{\rm eff}$ can be imposed by choosing external quark states 
with arbitrary kinematics.

The $C_i(\mu \approx M_{\rm SUSY})$ must then be evolved through RG flow to a scale
$\mu \sim$ few GeV. The anomalous dimension for the operator basis (\ref{basis})
needed in the RG equations is known to NLO accuracy \cite{NLOADM}. The amplitude for 
$B_d - \ov{B}_d$ mixing is finally obtained by multiplying the $C_i(\mu)$ by the 
matrix elements $\< Q_i \>(\mu)$ between the physical states of interest,
evaluated on the lattice \cite{DB-lattice}.

The leading order (LO) $C_i$ for $\D F =  2$ Hamiltonians in the strong sector
of the MSSM were first calculated in refs. \cite{GGMS}. They are simply obtained 
by computing four 
gluino-squark `box' diagrams (analogous to the $W$-quark boxes in the SM), and
rewriting their amplitude in terms of tree-level matrix elements of the
operators (\ref{basis}).

It is clear that knowledge of the next-to-leading order (NLO) corrections to the $C_i$ 
is the missing ingredient for a full NLO analysis of 
$\D F = 2$ processes. Such corrections are the subject of the present contribution.
In particular, section \ref{sec:NLO} briefly presents the highlights of the
calculation, that is dwelt on in a forthcoming publication. 
Section \ref{sec:pheno} addresses the phenomenological analysis, presenting 
preliminary results for the $B_d - \ov B_d$ system. A thorough analysis is again
postponed to an upcoming paper.

\vspace{-0.1cm}\section{NLO corrections to $\D F = 2$ Hamiltonians}\label{sec:NLO}

NLO corrections to the $C_i$ entering eq. (\ref{Heff}) are calculated by enforcing
the equality $\mc A_{\rm full} = \mc A_{\rm eff}$ at the scale $M_{\rm SUSY}$ up
to two loops in the full theory and correspondingly one loop in the effective
one. The two-loop diagrams mainly consist of gluon or squark corrections to the
LO boxes: an inventory taking into account multiplicities and `equivalences'
results in around $70$ diagrams. Hence, to be on the safe side as for the generation
and the initial treatment of the full amplitude, the {\Math} package {\FA} \cite{FA} 
was an essential tool.

To two-loop order it becomes important to choose a regularization scheme for
ultra-violet (UV) divergences as well as a kinematical configuration for the external 
quark legs. Concerning the first issue, we performed the whole calculation in
both the NDR and DRED ($\rm \ov{MS}$) schemes. As for external legs, we chose massless and 
zero-momentum quarks, since this leads to enormous simplifications in the algebra and
in the treatment of the two-loop integrals, that in this way are just ``vacuum''
ones. This choice brings about infra-red (IR) divergences that however must cancel in
the matching. In the intermediate steps IR divergences were coped with either by 
dimensional regularization (like UV ones) or by endowing gluon propagators with a mass 
term. 

We then compared the calculations in all the resulting UV-IR regularization schemes and 
obtained perfect consistency, taking into account the findings 
of \cite{ACMP} for the treatment of evanescents and of \cite{MV} for the fact that NDR 
breaks SUSY. Other important checks of the calculated $C_i$ were the already mentioned
disappearance of IR divergences and the fact that the $\mu$-dependence fulfills
the relevant RG equations. Explicit formulae will be reported elsewhere.

We note that the final results are in the so-called mass insertion approximation (MIA)
for the entries of the squark mass matrices in the `super-CKM' basis: within the
MIA diagonal entries are approximated with a common mass term $M_s^2$
appropriately chosen ($1^{\rm st}$ approx.) and the off-diagonal ones $\D_{XY}$
($X,Y$ being the four possible chiralities) are considered small with respect to 
$M_s^2$ and expanded as mass insertions $\de_{XY} = \D_{XY}/M_s^2 \ll 1$ ($2^{\rm nd}$
approx.). The MIA enormously simplifies the phenomenological analysis and in our
NLO case also decides for the {\em analytical} feasibility of the two-loop integrals.

\vspace{-0.1cm}\section{Phenomenological analysis of the $B_d - \ov B_d$ system}\label{sec:pheno}
The calculated NLO corrections to the $C_i$ at the matching scale, evolved to
the scale $\mu$ with the NLO ADM as reported in \cite{NLOADM}, permit for the
first time a full NLO control on $\D F = 2$ Hamiltonians (\ref{Heff}).
This allows a correspondingly accurate analysis of meson-antimeson oscillations,
updating the results of \cite{pheno}. We restrict here to the $B_d -\ov B_d$
sector and use the findings of ref. \cite{DB-lattice} for the lattice
determination of the matrix elements for the effective operators (\ref{basis}) 
in order to access the relevant amplitude 
$\mc A_{B_d} = \< B_d | (\mc H_{\rm eff}^{\D F = 2})^{\rm SM + SUSY}| \ov B_d\>$.

\begin{figure}[h!]
\vspace{-0.3cm}
\begin{center}
\includegraphics[scale=0.30]{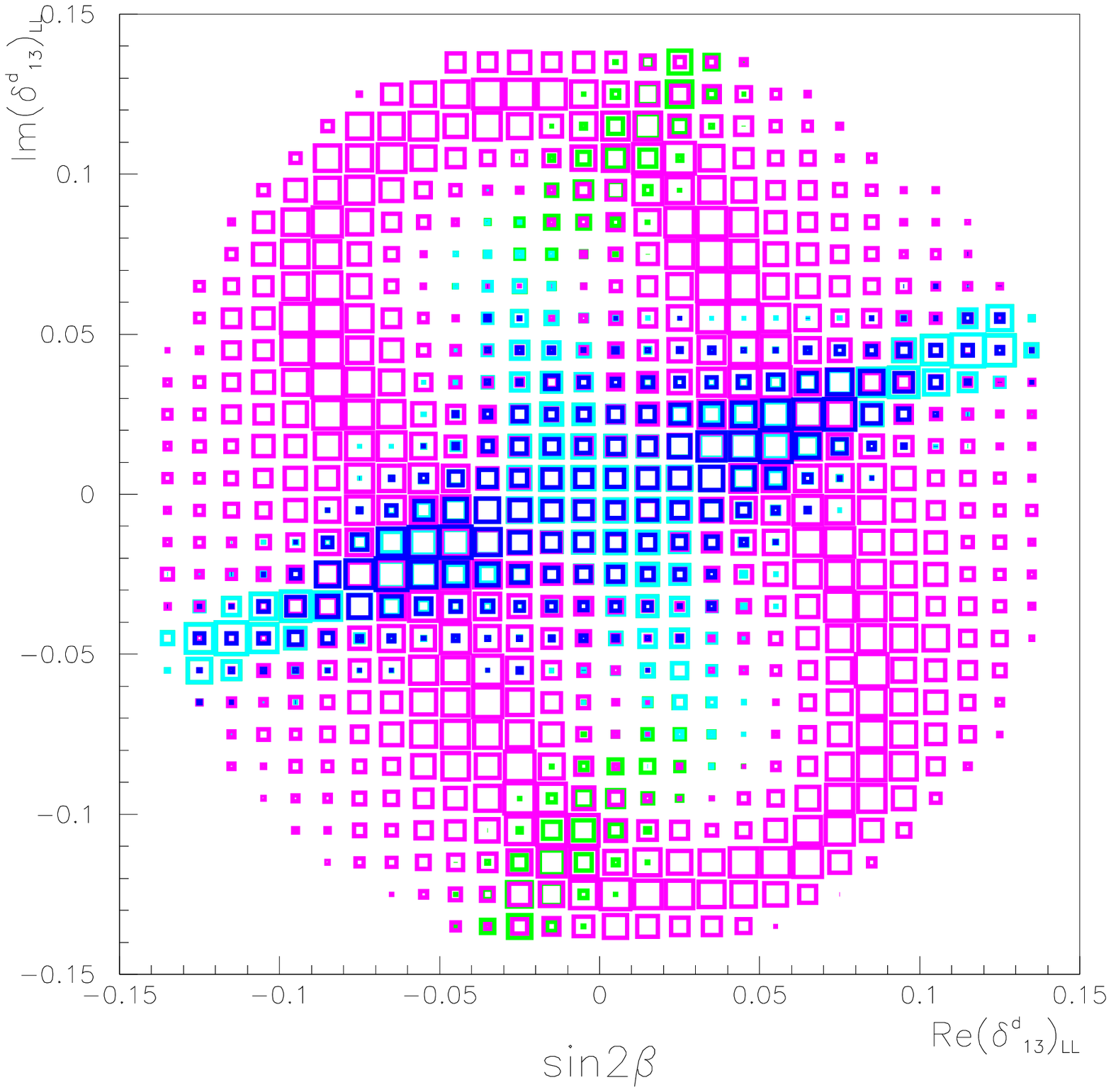}
\includegraphics[scale=0.6]{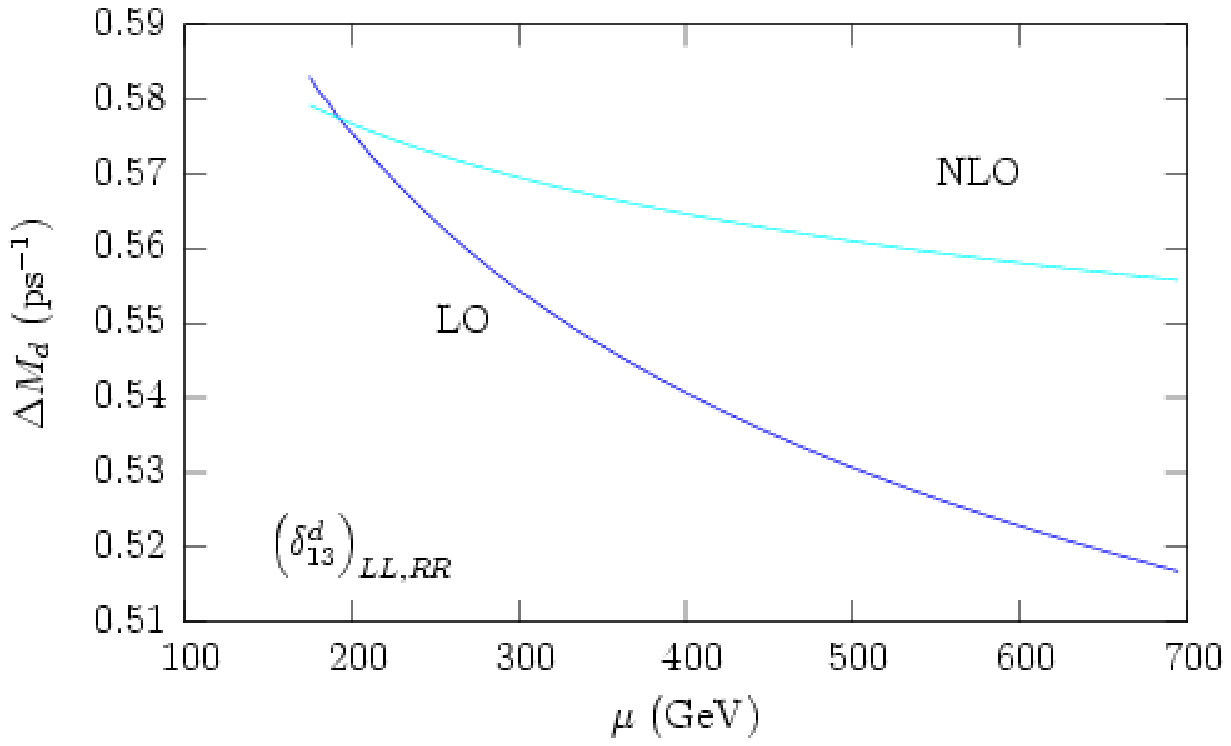}

\vspace{-0.5cm}
\hspace{-7.6cm}\colorbox[rgb]{1,1,1}{\phantom{XXX}}

\vspace{-0.4cm}
\caption{Left panel: allowed ranges for ${\rm Re}(\de^{13}_{LL})$ vs. ${\rm
Im}(\de^{13}_{LL})$. The sides of the `boxes' are proportional to the number of
`events' as weighted by the MonteCarlo. Constraints for different observables
are superimposed with different colors: $\D M_d$ only (magenta), $\sin(2 \be)$
only (green), $\sin (2 \be) \& \cos (2 \be)$ (cyan), all the constraints (blue).
Right panel: scale dependence of the $LL$ and/or $RR$ terms in the amplitude in the 
cases of Wilson coefficients to LO or to NLO accuracy (similar results hold in
the $LR$ and/or $RL$ case).}
\label{fig}
\end{center}
\vspace{-0.3cm}
\end{figure}

Our analysis proceeds as follows: one extracts with normal distributions around
their central experimental values all the (SM) parameters needed in $\mc
A_{B_d}$ (CKM entries are taken from tree-level processes only, to leave SUSY
unconstrained). The gluino mass is set to $M_{\tilde g} =$ 500 GeV and $x =
M^2_{\tilde g} / M_s^2 = 1$, while the mass insertions are extracted with flat
distributions according to the intervals: ${\rm Abs}(\de_{XY}) \in$ [0, 1] and 
${\rm Arg}(\de_{XY}) \in$ (-$\pi$, $\pi$]. Therewith one has MonteCarlo
determinations for $\mc A_{B_d}$, that are compared with experiments via the
observables $\D M_d = 2 {\rm Abs}(\mc A_{B_d})$ and 
$2 \be_{\rm eff} = {\rm Arg} (\mc A_{B_d})$, the latter being accessed by studying the 
decays $B_d \to J/\psi \, K_S^{(*)}$.
One can then place constraints on the mass insertions by `switching on' one 
$\de_{XY}$ at a time according to the choices: `$LL$ only', `$RR$ only', `$LL$ = $RR$', 
`$LR$ only', `$RL$ only', `$LR$ = $RL$'. Barring accidental cancellations among
amplitudes proportional to different $\de_{XY}$ parameters, the results can be interpreted 
as maximum ranges allowed by the present experimental information to the various mass insertions. 
Constraints on $\de_{LL}$ are displayed in fig. \ref{fig}, left panel, whereas the right panel
nicely shows the reduction in the scale dependence of $\mc A_{B_d}$ when only
$LL$ and $RR$ insertions are kept.

The `hierarchical' pattern of the constraints, \eg max($\de_{LL}$) $\gg$ 
max($\de_{RL}$), represents an {\em a posteriori} justification for the MIA, 
wherein possibly large interference effects among squark mass matrix parameters 
are not taken into account. Such pattern also motivates our separate
consideration of $LL, \, RR$ and respectively $LR, \, RL$ insertions in the
analysis of the scale dependence.

\vspace{-0.3cm}


\vspace{-0.8cm}
\begin{thebibliography}{99}
\bibitem{Alta}
G.~Altarelli,
CERN-PH-TH-2005-051.
\bibitem{GGMS}
J.~M.~Gerard, W.~Grimus, A.~Raychaudhuri and G.~Zoupanos,
Phys.\ Lett.\ B {\bf 140} (1984) 349.
J.~S.~Hagelin, S.~Kelley and T.~Tanaka,
Nucl.\ Phys.\ B {\bf 415} (1994) 293.~
F.~Gabbiani, E.~Gabrielli, A.~Masiero and L.~Silvestrini,
Nucl.\ Phys.\ B {\bf 477} (1996) 321.
\bibitem{NLOADM}
M.~Ciuchini, E.~Franco, V.~Lubicz, G.~Martinelli, I.~Scimemi and L.~Silvestrini,
Nucl.\ Phys.\ B {\bf 523} (1998) 501.~
A.~J.~Buras, M.~Misiak and J.~Urban,
Nucl.\ Phys.\ B {\bf 586} (2000) 397.
\bibitem{DB-lattice}
D.~Becirevic, V.~Gimenez, G.~Martinelli, M.~Papinutto and J.~Reyes,
JHEP {\bf 0204} (2002) 025.
\bibitem{FA}
T.~Hahn,
Comput.\ Phys.\ Commun.\  {\bf 140} (2001) 418.
\bibitem{ACMP}
G.~Altarelli, G.~Curci, G.~Martinelli and S.~Petrarca,
Nucl.\ Phys.\ B {\bf 187} (1981) 461.
\bibitem{MV}
S.~P.~Martin and M.~T.~Vaughn,
Phys.\ Lett.\ B {\bf 318} (1993) 331.
\bibitem{pheno}
M.~Ciuchini {\it et al.},
JHEP {\bf 9810} (1998) 008.~
D.~Becirevic {\it et al.},
Nucl.\ Phys.\ B {\bf 634} (2002) 105.

\end{thebibliography}
\end{document}